\documentstyle[12pt] {article}
\begin{document}

{\center {\Large QSO Pairs {\it across} Active Galaxies: Evidence of 
Blueshifts?\\}}

D. Basu {\it Department of Physics, Carleton University, Ottawa, ON K1S 
5B6, 
Canada
e-mail: basu@physics.carleton.ca\\}

{\bf Abstract.} Several QSO pairs have been reported  and their 
redshifts determined, where the two objects 
in each pair are located {\it across} an active galaxy. The usually 
accepted 
explanation of such occurrences is that the pair is ejected from the 
parent galaxy. Currently interpreted redshifted spectra for both the 
QSOs imply that both the objects are receding from the observer. 
However, 
ejection can occur towards and 
away from the observer with equal probability. We argue that for a 
system with two QSOs lying {\it across} the parent 
galaxy, ejection should have occurred in opposite directions, whereby 
one object will be approaching us and the other will be receding from 
us. The former would exhibit a blueshifted spectrum. We analyse here a 
sample of four such pairs and show that 
the observed spectrum of one QSO in each pair can be interpreted 
as blueshifted. The other exhibits the usual redshifted spectrum. A 
scenario 
based on the "sling-shot" mechanism of ejection is presented to explain 
the occurrences of the pairs in opposite sides of the active galaxies 
moving in opposite directions.\\ 


\pagebreak

\def\baselinestretch{1.66}
\large
\normalsize

{\center{\bf 1. Introduction\\}}

Literature search would reveal that there is a growing number of Quasi 
Stellar object (QSO) 
pairs being observed, where the two objects in the pair are located 
{\it across} 
an active 
galaxy. We present here a sample of four such pairs and their associated 
active 
galaxies (Table 1). In Table 1, column (1) gives the name of the active 
galaxy and the reference, column (2) gives its redshift, columns (3) and 
(5) give the names 
of the two QSOs forming the pair across the active galaxy, columns (4) and 
(6) give the 
redshifts of the two QSOs respectively. In each of these 
pairs, two QSOs have been 
identified and their redshifts determined, which are much larger than the 
redshift 
of the parent galaxy. Ejection from the parent galaxy is the explanation 
of such occurrences.

However, the basic mechanism of ejection necessitates that 
the process 
can occur in all directions with equal probability for randomly ejected 
objects. Objects being ejected may therefore move away from us as well as 
towards us, again, with equal probability. Such a situation is 
specially important and significant when the two objects are located 
{\it across} the parent galaxy, implying that the objects have been 
ejected by 
the galaxy to its two opposite sides, and hence, in two {\it opposite 
directions}. An ejection towards the observer would, of course, 
exhibit blueshifts in their observed spectra, in at least some cases 
under suitable conditions. In this respect, the kinematics of the ejection 
mechanism suggests that at {\it larger} redshifts ($\geq$ 3.5), "randomly 
directed motions will produce almost an order of magnitude more redshifts 
than blueshifts", and "As the mean speed of the sources increases, the 
fraction of blueshifts decreases" (Gordon 1980). This implies that for 
objects with 
{\it smaller} redshifts moving at slower speeds, the fraction of 
blueshifts to redshifts is higher than that for high redshift objects 
moving at larger speeds. All QSO pairs in our sample have redshifts 
smaller 
than 1.0, much smaller than the redshift of "$\geq$ 3.5" (Gordon 1980), 
and, hence, the 
probability of some of them being blueshifted is quite high. Furthermore, 
as commented by Popowski \& Weinzeirl (2003), "it is unrealistic to 
expect that the ejection process would always meet the conditions to 
produce only redshifts". As such, at least a certain fraction of the vast 
population of QSOs now known is expected to exhibit blueshifts. 

Nevertheless, the spectra of both 
objects in each pair have been interpreted as redshifted. Ejection has 
thus been considered away from us for {\it both} objects, without paying 
due consideration to the probability of the ejection towards the 
observer, apparently with the assumption that blueshifts are not 
observable in QSO spectra. The reality, however, is that no attempt is 
made in interpreting observed spectra in terms of blueshifts. The line 
identification program is such oriented as to ensure that all spectra are, 
almost as a rule, interpreted as redshifted. 

We suggest that 
the two objects in each QSO pair under investigation here have been 
ejected from the 
parent active galaxy in opposite directions, resulting in one approaching 
us 
and the other receding from us, which would result in  one 
of the objects exhibiting a blueshifted spectrum. The purpose of the 
present 
paper is to show that the spectrum of one 
object in 
each pair of the QSOs can indeed be interpreted as blueshifted.  

{\center{\bf 2. Blueshifted spectra\\}}

Putsil'nik (1979) demonstrated that the most frequently used UV lines 
CIII] 1909 and MgII 2798 for redshift identification of observed emission 
lines in QSOs may actually be misidentifications, and the IR lines of the 
Paschen series, viz. P$\beta$ 12818 and P$\alpha$ 18751 respectively, may 
be better identifications instead, producing "violet shifts". 
Additionally, literature search reveals many inconsistencies and 
inadequacies in redshift identifications. Thus, expected strong features 
are sometimes not seen at expected positions or observed as very weak 
lines or of unusual profiles, the redshift computed on the basis of 
observed lines may not be able to interprete additional features seen in 
subsequent observations, 
problems appear in energetics based on the calculated redshift values. We 
are therefore 
concerned about the correctness of the identification process, and, hence, 
about 
the possibility of misidentification of observed lines in the redshift 
determination of QSOs and other extragalactic objects. This 
necessitates alternative identifications and may lead to blueshifts.

It has been shown in recent years that blueshifts can interprete 
observed emission lines of QSOs, and, from various considerations, such 
interpretations in many cases are more convincing than the redshift 
interpretations (Basu \& Haque-Copilah 
2001). Very unusual spectra of three additional QSOs, viz. SDSS 1533-00, 
PG 1407+265 
and PKS 0635-752, emission and absorption, which cannot be explained 
properly by the redshift mechanism, have 
been interpreted in terms of blueshifts (Basu 2004). Spectra of two active 
galactic nuclei (AGN), viz. PKS 
2149-306 and CXOCDFS J033225.3-274219, each exhibiting an X-ray emission 
feature, which cannot be explained by the redshift determined from its 
optical spectrum, have been 
successfully interpreted in terms of blueshifts (Basu (2006a).  
Blueshfts 
have also been demonstrated to explain observed spectra of other 
extragalactic objects, viz. several high redshift galaxies (Basu 1998),
the galaxy STIS123627+621755 (the redshift interpretation being unable to 
explain this spectrum) (Basu 2001a), host galaxies of 
several supernovae Ia (Basu 2000), host galaxies of several gamma ray 
bursts (Basu 2001b), and also the puzzling spectra of the galactic X-ray 
source 1E 1207.4-5209 associated with the SNR G296.5+10.0 (Basu 
2006b). In all the above cases, serious inconsistencies exist 
in redshift interpretations of the spectra of the objects concerned. It is 
therefore possible that redshifts have been assigned 
to spectra of some extragalactic objects due to misidentification of 
observed lines.

{\center{\bf 3. Results\\}}

We re-examined the observed spectra of the eight QSOs of the four pairs 
in our sample, 
and found several incompleteness and inadequacies in the redshift 
identification process of some of the objects, as discussed in Sec. 
4. As such, we have interpreted the spectra of one object 
in each pair as 
blueshifted by identifying the observed lines with search lines of longer 
wavelengths, and have determined the 
blueshift 
values of each line and, hence, of each such object. We have followed the 
standard procedure of the identification process for the blueshift 
determination, viz. at least two 
observed lines must exhibit the same 'shift' (red or blue), whether 
emission or absorption, when identified 
with two separate search lines, and the same 'shift' must be 
exhibited for any 
third or more lines (Basu 1973a, 1973b). Furthermore, when 
necessary, the higher order line(s) of a series (Balmer or Paschen) has 
been identified, the lower order line(s) being outside the 
observed region of the spectrum.
Also, in some cases, lower order line(s) of a series has been 
identified, the higher order line(s) may be too weak to be 
seen.  Results are shown in Table 2. 

In Table 2, column (1) is the name of the active galaxy, column 
(2) gives the name of the QSO, whether the features are emission (EM) or 
absorption (ABS) and the reference of its spectrum, column (3) is the 
observed wavelength of 
the line in the spectrum ($\lambda_{o}$), column (4) is the wavelength 
of the search line 
identified with the observed line for determining the redshift 
($\lambda_{r}$), column (5) 
is the redshift value of the line (z$_{r}$), column (6) is spread in 
redshift 
($\Delta$z$_{r}$) measured as the difference between the 
maximum and minimum values of redshifts in the system, column (7) is the 
wavelength of the search line identified with the observed 
line for determining the blueshift ($\lambda_{b}$), column (8) is the 
blueshift value of 
the observed line (z$_{b}$), column (9) is the spread in blueshift 
($\Delta$z$_{b}$) 
measured as the difference between the maximum and minimum values of 
blueshifts in the system.

The mean blueshifts of the four objects, from Table 2, are 0.7065, 
0.7502, 0.0566 and 0.2340, respectively from top downwards. Additionally, 
two of the QSOs, viz. 1218+75 and 1141+20 (QSO2), exhibit absorption lines 
as well, and these have been identified with search lines of longer 
wavelengths to obtain mean blueshifts 0.7890 and 0.3450 for the two 
objects respectively. 

Absorption lines are known to be formed  partly by hydrogen clouds at 
various stages of evolution to galaxy formation and partly by metals 
produced in haloes of galaxies.

{\center{\bf 4. Discussion\\}}

We have identified the two emission lines at 3720\AA and 5565\AA in the 
QSO 1120+13 with the IR lines P$\beta$ 12818 and P$\alpha$ 18751, instead 
of the UV lines CIII] 1909 and MgII 2798 respectively. The possibility 
that these two IR lines are better 
identifications of some observed lines rather than the UV lines was 
suggested by Putsil'nik (1979), as mentioned earlier (Sec. 2).

The QSO 1416+25 exhibits only one emission line, viz. that at 4684\AA, 
which has been identified with MgII 2798 in the redshift interpretation. 
In principle, it can be identified with any search line, and the proper 
identification needs at least two lines, as explained above. However, in 
the absence of any other 
line and 
any other information of this line, and following Putsil'nik (1979), we 
have 
identified the line with P$\alpha$ 18751. Other possible alternative is, 
of course, 
H$\alpha$ 6563, but that would require a much stronger line. On the other 
hand, H$\beta$ 4861 is ruled out in this case as that would bring 
H$\alpha$ 6563 within the observed region which is not seen. 

For the QSO 1218+75, the blueshift identification of the line at 4600\AA 
with H$\beta$ 4681 is justified as the H$\alpha$ 6563 expected within the 
observing region is seen at 6173\AA. Both are fairly strong lines. The 
absorption lines are identified in the blueshift interpretation with 
four of the strongest molecular hydrogen (H2) lines. 

The emission lines of the QSO 1141+20 (QSO 2) have been shown earlier to 
be better interpreted as blueshifted (Basu \& Haque-Copilah 2001). But the 
spectrum exhibits absorption lines also, and here 
we treat the complete spectrum - both emission and absorption. On the 
other hand, the redshift interpretation of the spectrum of 1141+20 (QSO 2) 
is very confusing. It has been 
described as "difficult to identify" for the redshift measurement, with 
"unusual features", and "H$\gamma$ and H$\alpha$ in emission, if present 
at all, are very weak". Two emission 
lines at 4050\AA and 4494\AA, and the broad "so strong" absorption line at 
4300\AA cannot be identified in the redshift interpretation. In addition, 
the 
absorption line at 3528\AA has 
been identified with a multiple of lines viz. SiII 1808-1817. We have 
identified {\it all} the observed lines in this object in
the blueshift interpretation - the above two emission lines as [CaV] 
5301 and HeI 5876 respectively, and the broad absorption line as the 
H$\alpha$ 6563, the strongest of the search lines.

It should be noted here that the blueshift values presented in this paper 
are results of superposition of the cosmological redshifts and the 
Doppler shifts due to the object approaching the observer.

{\center{\bf 5. The 'spread' statistic: goodness of fit\\}}

The quantity spread, $\Delta$z in Table 2 (columns (6) and (9)), is a 
measure of the goodness of fit for the identification of the observed 
lines with search lines for computing the 'shift', red or blue, 
for a system. In principle, $\Delta$z should be as close to zero as 
possible. However, in practice, the quantity depends on, and is very 
sensitive to, the value of the observed wavelength ($\lambda_{o}$,
column (3), Table 
2) which is used to evaluate the 'shift'. Unfortunately, the exact 
determination of $\lambda_{o}$ is very difficult in practice, even in 
high 
s/n and high resolution spectra, as the line profile may be broad, 
double- or multi-peaked, with complicated structure, blending, etc. Physically, 
these may involve net flows, screening, gradients, partial absorption, 
etc. Local velocity drifts, infall or outflow of materials (yet undecided, 
Penston et al. 1990), etc. may 
also move 
$\lambda_{o}$ away from the expected position. Uncertainties 
in the determination of $\lambda_{o}$ due to the above reasons may, in 
some cases, result in somewhat larger values of $\Delta$z, in both 
redshifts and blueshifts. At least upto certain extent, this should 
not be considered as errors in 
measurement, but may have physical reasons. Having said that, Table 1 
would show that, in general, $\Delta$z$_{b}$, i.e. 
spreads in blueshift values, are somewhat smaller, and hence closer to 
zero, than 
$\Delta$z$_{r}$, i.e. 
spreads in redshift values. 

{\center{\bf 6.  A possible scenario: ejection mechanism\\}}

The analysis presented here shows that QSO pairs {\it across} active 
galaxies comprise two objects - one approaching us and thus exhibiting 
blueshifts, the other moving away from us and thus exhibiting redshifts. 
Such a scenario can be achieved in terms of the  
ejection mechanism. Merger of black holes is known to lead to their 
ejections in opposite directions by the so-called "sling-shot" mechanism 
when the system becomes unstable (Saslaw et al. 1974; Valtonen 1976a, 
1976b). Merging of galaxies, each hosting a 
supermassive black hole at its 
centre, which are known 
to be seats of activities (Basu et al. 1993; 
Capetti et al. 2005), may produce such a situation. Initially, a binary 
system is formed by the two 
central black holes (Valtoja et al. 1989), and such a system has been 
reported to be detected in NGC 6240 (Kommossa et al. 2003). As the merger 
process proceeds further, the supermassive black holes (primaries) are 
ejected at relavistic or non-relativistic speeds "in two opposite 
directions" (Mikkola \& Valtonen 
1990). Evidence of ejection of a supermassive black hole by the "sling 
shot" mechanism resulting from merger of galaxies has recently been 
presented (Haehnelt et al. 2005). Additionally, 
satallite black holes of intermediate masses are also believed to 
accompany the central supermassive black holes in galaxies (Carr 1978; 
Carr et al. 1984) and are also ejected, some of them assuming eccentric 
orbits 
around the primaries (Valtonen \& Basu 1991). 

It is further known that a black hole at the centre of a galaxy possesses 
a gaseous accretion disk around it, and this is believed to survive the 
tidal disruption accompanying the ejection process (Rees \& Saslaw 1975; 
Lin \& Saslaw 1977; De Young 1977). A QSO may be formed by the 
interaction between the disk with the black hole and the surrounding (Rees 
1984; Osterbrock \& Mathews 1986; Valtonen \& Basu 1991; Spriegel 2005). 
One can further envisage that the same process of interaction would occur 
between the 
surroundings and the gaseous disks around the satellite black holes, 
as in case of their primary counterparts, although at reduced scales as
their masses are smaller, and eventually some of the satallite black holes 
may end up as faint or 
nascent or other galaxies. 

The merger of two galaxies thus finally gives birth to two QSOs 
ejected in 
opposite directions. Each QSO may be accompanied by one or more 
galaxy-like objects, and 
the latter act as absorbing clouds when aligned along the line of sight. 
It may be noted in this connection that absorbing clouds have been shown 
earlier to be linked with the birth of QSOs themselves (Basu 1982). 
Moreover, faint, nascent and other galaxies asociated with QSO-like 
objects have 
been reported to be observed (Dressler et al. 1993; Tripp et al. 1998; 
Tresse et al. 1999).

It can 
therefore be conceived that QSO pairs {\it across} active galaxies are the 
results of the ejection process, viz. the "sling shot" mechanism. One of 
the pairs is approaching us 
exhibiting blueshifts in emission  and absorption lines, the latter, when 
observed, being 
produced by the accompanying absorbing clouds in the form of faint or 
nascent or other galaxies ejected by the same process but at larger speeds 
and 
hence showing larger 
absorption blueshifts than the corresponding emission blueshifts. The 
other is receding from us exhibiting the usual redshifted spectra.

{\center{\bf 7. Concluding remarks\\}}

We have followed an entirely different path, viz. blueshifts in 
extragalactic spectra, and not a variant of existing ones, to interprete 
observational data. We are motivated by the possibility that blueshifts 
have been ignored  by the astronomical community in the modern line 
identification programs. This, in its turn, has led to wrong approaches. 
One such example is the assmption that the ejection mechanism would always 
occur away from the observer, thus obeying all the rules of redshifts 
only. We have shown here that blueshifts are real possibilities, and 
observations obey the basic notion of the ejection process, viz. objects 
should be ejected in all directions with equal probability. 

Finally, it should be emphasized that blueshifts and redshifts are 
complementary and not contradictory. Modern observations would bring in 
many unusual systems some of which may not be explained by the usual 
redshift hypothesis, and blueshifts may explain them. Ignoring blueshifts 
may result in missing some important cosmological scenarios. 

{\center{\bf Acknowledgement}}

The author thanks an anonymous referee for helpful suggestions.

\def\baselinestretch{1.00}
\large
\normalsize

\pagebreak

{\center{\bf References\\}}

\begin{description}

\item Arp, H., 1984, {\it Astrophys. J.} 283, 59 
\item Arp, H., 1995, {\it Astron. Astrophys.} 316, 57
\item Arp. H., 1997, {\it Astron. Astrophys.} 319, 33
\item Arp, H., et al., 2002, {\it Astron. Astrophys.} 391, 833
\item Basu, D., 1973a, {\it Nat. Phys. Sci.} 241, 159
\item Basu D., 1973b, {\it The Observatory} 93, 229
\item Basu, D., 1982, {\it Ap. Letts.} 22, 139 
\item Basu, D., 1998, {\it A \& SS} 259, 415
\item Basu, D., 2000, {\it Mod. Phys. Letts.} A 15, 2357
\item Basu, D., 2001a, {\it Ap. Letts. \& Comm.} 40, 157
\item Basu, D., 2001b, {\it Ap. Letts. \& Comm.} 40, 225
\item Basu, D., 2004, {\it Phys. Scr.} 69, 427
\item Basu, D., 2006a, {\it Astr. J.} 131, 1231
\item Basu, D., 2006b, {\it Astr. Nachr.}, 327, 724
\item Basu, D. \& Haque-Copilah, S., 2001, {\it Phys. Scr.} 63, 425
\item Basu, D., et al. 1993, {\it Astron. Astrophys.}, 272, 417
\item Capetti, A., et al, 2005, {\it Astron. Astrophys.} 431, 465
\item Carr, B., 1978, {\it Comm. Astrophys.} 7, 161
\item Carr, B., et al., 1984, {\it Astrophys. J.} 277, 445
\item De Young, D., 1977, {\it Astrophys. J.} 211, 329
\item Dressler et al., 1993, {\it Astrophys. J.}, 404, L45
\item Gioia, I,. 1984, {\it Astrophys. J.} 283, 495 
\item Gordon, K., 1980, {\it Amer. J. Phys.} 48, 524
\item Haehnelt, M. G., et al., 2006, {\it Mon. Not. R. Astr. Soc.}, 366, 
L22
\item Kommossa, S., et al., 2003, {\it Astrophys. J.} 592, L15 
\item Lin, D. \& Saslaw, W., 1977, {\it Astrophys. J.} 217, 958
\item Mikkola, S. \& Valtonen, 1990, {\it Astrophys. J.} 348, 412
\item Osterbrock, P. \& Mathews, W., 1986, {\it ARAA}, 24, 171 
\item Penston, M. V., et al., 1990, {\it Mon. Not. R. Astr, Soc.} 244, 357
\item Popowski, P. \& Weinzierl, W., 2004, {\it Mon. Not. R. Astr. Soc.} 
348, 235 
\item Putsil'nik, A., 1979, {\it Astron. Astrophys.} 78, 248
\item Rees, M., 1984, {\it ARAA} 22, 471
\item Rees, M. \& Saslaw, W., 1975, {\it Mon. Not. R. Astr. Soc.} 171, 53 
\item Saslaw, W., et al. 1974, {\it Astrophys. J.} 190, 253 
\item Spriegal, V., et al., 2005, {\it Astrophys. J.} 620, L79
\item Stocke, J., et al., 1983, {\it Astrophys. J.} 273, 458
\item Teresse, L., et al, 1999, {\it Astron. Astrophys.} 346, L21
\item Tripp, T., et al., 1998, {\it Astrophys. J.} 508, 200
\item Valtoja, E., et al., 1989, {\it Astrophys. J.} 343, 47
\item Valtonen, M., 1976a, {\it Astron. Astrophys.} 46, 429
\item Valtonen, M., 1976b, {\it Astron. Astrophyhs.} 46, 435
\item Valtonen, M. \& Basu, D., 1991, {\it J. Astrophys. Astr.} 12, 91

\end{description}

\pagebreak

\begin{tabular}{llllllr}
\multicolumn{6}{c}{{\bf Table 1.} QSOs across galaxies and redshifts} \\ 
\hline
\multicolumn{1}{c}{Galaxy (1)}
& \multicolumn{1}{c}{z$_{r}$ (2)}
& \multicolumn{1}{c}{QSO (3)}
& \multicolumn{1}{c}{z$_{r}$ (4)} 
& \multicolumn{1}{c}{QSO (5)}
& \multicolumn{1}{c}{z$_{r}$ (6)} \\ \hline

NGC 3628(i) & 0.0028 & 1WGAJ 1120.2+1332 & 0.995 & 1WGAJ 1120.4+1340 & 
0.981 \\
NGC 3842(ii) & 0.0260 & 1141.8+2013 (QSO 1) & 0.335 & 1141.4+2014 (QSO 2) 
& 0.946 \\
NGC 5548(iii) & 0.0170 & 1E 1415.1+2527 & 0.560 & 1E 1416.7+2526 & 0.674 
\\
Mark 205(iv) & 0.0700 & 1218.7+7522 & 0.645 & 1219+7533 & 0.460 \\ \hline 

\end{tabular}

References: (i) Arp et al. (2002); (ii) Arp (1984); (iii) Arp (1997); (iv) 
Arp (1995).

\pagebreak

\begin{tabular}{lllllllllr}
\multicolumn{9}{c}{{\bf Table 2.} Redshifts and blueshifts in observed 
lines of five QSOs} \\ \hline
\multicolumn{1}{c}{Galaxy (1)}
& \multicolumn{1}{c}{QSO (2)}
& \multicolumn{1}{c}{$\lambda_{o}$ (3)}
& \multicolumn{1}{c}{$\lambda_{r}$ (4)}
& \multicolumn{1}{c}{z$_{r}$ (5)}
& \multicolumn{1}{c}{$\Delta$z$_{r}$ (6)}
& \multicolumn{1}{c}{$\lambda_{b}$ (7)}
& \multicolumn{1}{c}{z$_{b}$ (8)} 
& \multicolumn{1}{c}{$\Delta$z$_{b}$ (9)} \\ \hline

NGC 3628 & 1120+13 & 3720 & CIII] 1909 & 0.9800 & 0.008 & P$\beta$ 12818 
& 0.7098 & 0.0060 \\
          & EM (i) & 5565 & MgII 2798 & 0.9880 & & P$\alpha$ 18751 & 
0.7032 \\ \hline
NGC 5548 & 1416+25 & 4684 & MgII 2798 & 0.6740 & & P$\alpha$ 18751 & 
0.7502 \\ 
         & EM (ii) \\ \hline   
Mark 205 & 1218+75 & 4600 & MgII 2798 & 0.6440 & 0.012 & H$\beta$ 4861 
& 0.0537 & 0.0060 \\
         & EM (iii) & 6173 & [OII] 3727 &0.6564 & & H$\alpha$ 6563 & 
0.0594 \\ \hline
         & ABS (iii) & 4275 & FeII 2587 & 0.6525 & 0.011 & H2 20338 & 
0.7898 & 0.0087 \\
         & & 4307 & FeII 2600 & 0.6565 & & H2 20587 & 0.7908  \\
         & & 4600 & MgII 2796 & 0.6452 & & H2 21218 & 0.7832 \\ 
         & & 4627 & MgII 2803 & 0.6507 & & H2 22233 & 0.7919 \\ \hline
NGC 3842 & 1141+20 & 3714 & CIII] 1909 & 0.9460 & 0.003 & H$\beta$ 4861 
& 0.2360 & 0.0100 \\ 
         & (QSO 2) & 4050 & ? & ? & & [CaV]5301 & 0.2370 \\
         & EM (iv) & 4494 & ? & ? & & HeI 5876 & 0.2350 \\
         & & 4804 & [OII] 2470 & 0.9450 & & [OI] 6300 & 0.2375 \\
         & & 5444 & MgII 2798 & 0.9457 & & HeI 7065 & 0.2294 \\
         & & 8455 & H$\gamma$ 4340 & 0.9482 & & P$\gamma$ 10938 & 0.2270 
\\ \hline
         & ABS (iv) & 3528 & SiII 1808- & 0.9510 & 0.023 & SiII 5454 & 
0.3531 & 0.0183 \\
         & & & 1817 & 0.9420 & 0.019 \\
         & & 4300? & ? & ? & & H$\alpha$ 6563 & 0.3348 \\
         & & 8369? & H$\gamma$ 4340? & 0.9280 & & P$\beta$ 12818 & 0.3471 
\\ \hline 

\end{tabular}

References: (i) Arp et al. (2002); (ii) Stocke et al. (1983); (iii) Gioia 
et al. (1984); (iv) Arp (1984) (see also Basu \& Haque-Copilah 2001).

\end{document}